\documentclass[a4paper]{JAC2003}
\usepackage{graphicx}
\usepackage{booktabs}
\usepackage[normalem]{ulem}
\usepackage{subfigure,tikz}
\usepackage{lineno}
\usepackage{tikz}
\usetikzlibrary{trees}
\usepackage{color}
\usepackage{xcolor}
\begin{document}

\title{BEAM--BEAM EFFECTS IN THE HIGH-PILE-UP TESTS OF THE LHC}
\author{G. Trad, CERN, Geneva, Switzerland}
\maketitle

\begin{abstract}
	Investigating the beam--beam limit in the LHC is of great importance, since identifying its source is crucial for the luminosity optimization scenario. Several experiments were carried out to search for this limit and check whether it is dominated by the head-on (HO) or the long-range (LR) interactions. In this paper only the HO collision effects will be considered, tracking the evolution of the maximum tune shift achieved during the dedicated machine developments and the special high pile-up fills.
\end{abstract}

\section{INTRODUCTION}
Contrarily to electron machines, hadron machines are mostly limited by non-linear effects and lifetime problems, in particular for beams with high intensities~\cite{IPAC_2011}. The limitation appears as a slow emittance increase (over hours) or beam losses (tails and dynamic aperture), bad lifetime or other effects such as coherent beam--beam oscillations.

It is important to note that the head-on (HO) tune shift depends only on the bunch intensity and the normalized emittance, i.e. is independent of the beta function at the interaction point $ \beta^{*} $ and the energy as shown in Eq.~(\ref{eq:HO_tune_shift}). For the long-range (LR) interactions, the tune shift depends on the beam separation $d_{\mathrm{sep}}$, thus on the crossing angle $\alpha$ and the beam energy $\gamma$, as shown in Eq.~(\ref{eq:LR_tune_shift}).

\begin{equation}
\Delta  Q_{\mathrm{HO}} \propto \frac{N}{\varepsilon_{n}},
\label{eq:HO_tune_shift}
\end{equation}

\begin{equation}
L \propto \frac{N^{2}}{\varepsilon_{n}} = \Delta  Q_{\mathrm{HO}} \cdot N,
\label{eq:Luminosity}
\end{equation}

\begin{equation}
\Delta  Q_{\mathrm{LR}} \propto \frac{N}{d_{\mathrm{sep}}^2} = \frac{N\cdot\varepsilon_{n}}{\alpha^2 \cdot \beta^{*} \cdot \gamma}.
\label{eq:LR_tune_shift}
\end{equation}

Therefore, identifying the source of these limits has vital significance for the luminosity optimization strategy:
if the HO collisions are dominating the beam--beam limit, it is then advantageous to increase the bunch intensity together with the transverse emittance since this would keep the tune shift unaffected and increases the luminosity proportionally to the intensity (see Eq.~(\ref{eq:Luminosity})). The luminosity is further increased by pushing more the focusing at the interaction point (IP) without affecting the beam--beam parameter $ \xi $:

\begin{equation}
\xi_{21x,y}=\frac{N_{1}r_{p}}{2\pi\sqrt{\varepsilon_{x,y_{n1}}}\left(\sqrt{\varepsilon_{x,y_{n1}}} + \sqrt{\varepsilon_{y,x_{n1}}}\right)}.
\label{eq:XI_definition}
\end{equation}

On the other hand, when the machine is limited by the LR interactions, a large number of bunches with a moderate  $ \beta^{*} $ is preferred.
Since $\Delta  Q_{LR}$ depends on  $ \beta^{*} $, any change of the optical function at the IP requires one to adjust the crossing angle $ \alpha $ to keep the LR tune shift constant,
\begin{equation}
\alpha \propto \sqrt{\frac{N\cdot\varepsilon_{n}}{\Delta  Q_{\mathrm{max}} \cdot \beta^{*} \cdot \gamma}}.
\label{eq:Xing_Angle_adjust_for_LR}
\end{equation}

Some confusion is related to the maximum expected HO beam--beam tune shift for the LHC. The nominal HO tune shift was derived from the experience of the Super Proton Synchrotron (SPS), taking into account possible contributions from the lattice non-linearities and significant LR contributions. The nominal value of $\xi  = 0.0037$ was defined to provide a coherent set of parameters to reach the target luminosity, $L=10^{34}$\,cm$^{-2}$\,s$^{-1}$~\cite{LHC_DESIGN_REPORT, LHC_bb_review_and_outlook}.
Therefore, it should be considered as a conservative value and not as a real upper limit, in particular in the absence of strong LR interactions.

\section{Is the `nominal' beam--beam HO tune shift reachable?}
Early in 2010, even if no show-stoppers were expected, dedicated experiments and observations during normal operation were planned to check the feasibility of colliding beams with the nominal tune shift estimated in the design report.
In the following, we briefly list a few of these fills including the important results observed~\cite{Performance_Note_2010_39}.

\subsection{{Fill 1069 (May 2010)}}

\subsubsection{Experimental setup}

\tikzstyle{every node}=[draw=black,thick,anchor=west]
\tikzstyle{selected}=[draw=red,fill=red!30]
\tikzstyle{optional}=[dashed,fill=gray!50]
\begin{figure}[!h]
\centering
\begingroup
    \fontsize{8pt}{8pt}\selectfont
{
\begin{tikzpicture}[scale=1,%
  grow via three points={one child at (0.2, --0.7) and
  two children at (0.2, --0.7) and (0.2, --1.4)},
  edge from parent path={(\tikzparentnode.south) |- (\tikzchildnode.west)}]
  \node {Beam 1}
    child { node {\color{blue} B1-b1}
      child { node {IP1 vs. \color{red} B2-b1}}
      child { node {IP5 vs. \color{red} B2-b1}}
      child { node {IP8 vs. \color{red} B2-b892}}
    }
    child [missing] {}				
    child [missing] {}				
    child [missing] {}		
    child { node {\color{blue} B1-b1786}
      child { node {IP8 vs. \color{red} B2-b1}}
    }
    child [missing] {};
\end{tikzpicture}
}
{
\begin{tikzpicture}[scale=1,%
  grow via three points={one child at (0.2, --0.7) and
  two children at (0.2, --0.7) and (0.2, --1.4)},
  edge from parent path={(\tikzparentnode.south) |- (\tikzchildnode.west)}]
  \node {Beam 2}
    child { node {\color{red} B2-b1}
      child { node {IP1 vs. \color{blue} B1-b1}}
      child { node {IP5 vs. \color{blue} B1-b1}}
    }
    child [missing] {}				
	child [missing] {}		
    child { node {\color{red} B2-b892}
      child { node {IP2 vs. \color{blue} B1-b1}}
      child { node {IP8 vs. \color{blue} B1-b1786}}
    }
    child [missing] {};
    child [missing] {};
\end{tikzpicture}
}
\endgroup
\caption{Collision scheme of the injected bunches in Fill 1069. Shown are the colliding bunch pairs in the two beams in the four interaction points. The slot numbers are given to specify a bunch position.}
\label{fig:Collision Scheme of the bunches in fill 1069}
\end{figure}
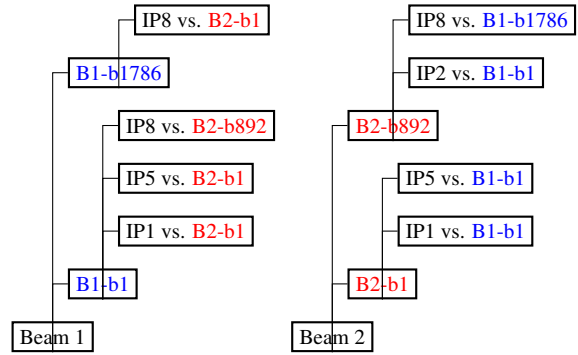

For the LHC fill in study (Fill 1069), two `almost' nominal bunches of intensity $\sim$1$\times$\,$10^{11}$\,ppb and normalized emittance of 3\,$\mu$m were injected in each ring. \\
The bunches were put in collision according to the standard operational cycle, with $\beta* = 10$\,m in all IPs and the collision tunes ($Q_{x} = 0.31$, $Q_{y} = 0.32$), at the injection energy of 450\,GeV. The collision scheme is shown in Fig.~\ref{fig:Collision Scheme of the bunches in fill 1069}. \\

\begin{figure}[]
\centering
\includegraphics[width=.48\textwidth,angle=-00]{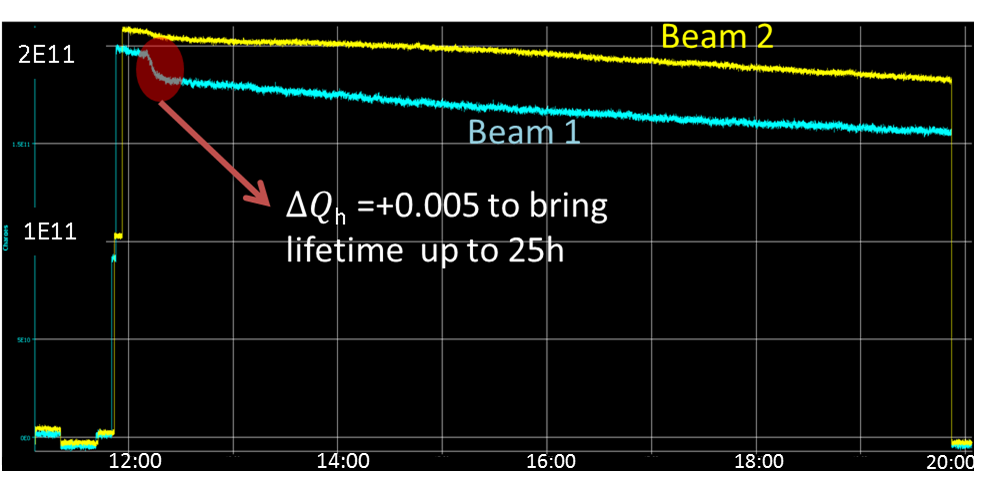}
\caption{Total beam intensity evolution during Fill 1069. Beam 1 is shown in blue while Beam 2 is shown in yellow.}
\label{fig:BB_Wshop_1068_Intensity}
\end{figure}

\subsubsection{Observations}
When the separation bumps were `collapsed' simultaneously to bring the beams into collision, the lifetime dropped, especially in Beam 1, but a small tune adjustment ($\Delta Q_{h} \sim +0.005$) was sufficient to stabilize it at 25\,h (Fig.~\ref{fig:BB_Wshop_1068_Intensity}).
The resulting beam--beam tune shift exceeded the nominal value reaching $\xi = 0.004$ per IP. An increase of the vertical  emittance in Beam 2 was observable and thought to be originating from an external excitation not fully understood (the hump).
Since the results were encouraging, and no limitations were expected for the HO tune shift, it was decided to increase the intensity of the colliding bunches exploring regions with a higher tune shift.

\subsection{{Fill 1765 (May 2011)}}

\subsubsection{Experimental setup}
One high-intensity bunch was injected per beam ($\sim$1.6$\times$\,$10^{11}$\,ppb); the measured normalized emittance by the wire scanners (WSs) was $\varepsilon _{_{x,y}}=1.2$\,$\mu$m.
With the collision tunes setting, the beams were brought into collision at injection energy. The IP settings were the following: $\beta* = 11$\,m and nominal crossing angles in all IPs, while the spectrometers were off in IP2 and IP8.
The transverse damper (ADT) was turned on only at injection and set off afterwards~\cite{MD_Note_2011_29,MD_Note_2011_58}.

\subsubsection{Observations}
An increase in the vertical emittance once in collision was observed, resulting in a 2.2\,$\mu$m emittance and $\xi=0.009$/IP. Since the bunches were colliding only in IP1 and IP5, a total  $\xi=0.018$ was reached.
A small tune scan was tried as well at the end of the fill, to search for a better working point. No lifetime effects were observed, just a minor emittance blow-up for Beam 2.

\subsection{{Fill 1766: Part 1 (May 2011)}}
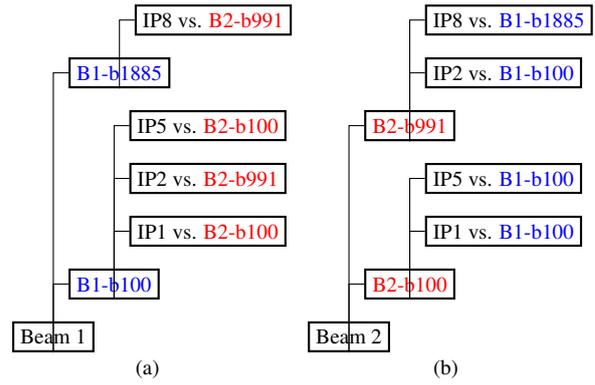
\begin{figure}
\centering
\begingroup
    \fontsize{8pt}{8pt}\selectfont
\subfigure[]
{
\begin{tikzpicture}[scale=1,%
  grow via three points={one child at (0.2, --0.7) and
  two children at (0.2, --0.7) and (0.2, --1.4)},
  edge from parent path={(\tikzparentnode.south) |- (\tikzchildnode.west)}]
  \node {Beam 1}
    child { node {\color{blue} B1-b100}
      child { node {IP1 vs. \color{red} B2-b100}}
      child { node {IP2 vs. \color{red} B2-b991}}
      child { node {IP5 vs. \color{red} B2-b100}}
    }
    child [missing] {}				
    child [missing] {}				
    child [missing] {}		
    child { node {\color{blue} B1-b1885}
      child { node {IP8 vs. \color{red} B2-b991}}
    }
    child [missing] {};
\end{tikzpicture}
}
\subfigure[]
{
\begin{tikzpicture}[scale=1,%
  grow via three points={one child at (0.2, --0.7) and
  two children at (0.2, --0.7) and (0.2, --1.4)},
  edge from parent path={(\tikzparentnode.south) |- (\tikzchildnode.west)}]
  \node {Beam 2}
    child { node {\color{red} B2-b100}
      child { node {IP1 vs. \color{blue} B1-b100}}
      child { node {IP5 vs. \color{blue} B1-b100}}
    }
    child [missing] {}				
	child [missing] {}		
    child { node {\color{red} B2-b991}
      child { node {IP2 vs. \color{blue} B1-b100}}
      child { node {IP8 vs. \color{blue} B1-b1885}}
    }
    child [missing] {};
    child [missing] {};
\end{tikzpicture}
}
\endgroup
\caption{Collision scheme of the injected bunches in Fill 1766.}
\label{fig:Collision scheme of the bunches in fill 1766 p1}
\end{figure}

Following the encouraging beam--beam parameter reached in Fill 1765, it was decided to maintain the high-intensity bunches and increase the number of collisions. Therefore, two bunches per beam were injected, allowing collisions in all interaction points but with a different collision pattern, as shown in Fig.~\ref{fig:Collision scheme of the bunches in fill 1766 p1}.

\tikzstyle{every node}=[draw=black,thick,anchor=west]
\tikzstyle{selected}=[draw=red,fill=red!30]
\tikzstyle{optional}=[dashed,fill=gray!50]


\subsubsection{Observations}

\begin{figure}[h!]
\centering
\includegraphics[width=.48\textwidth,angle=-00]{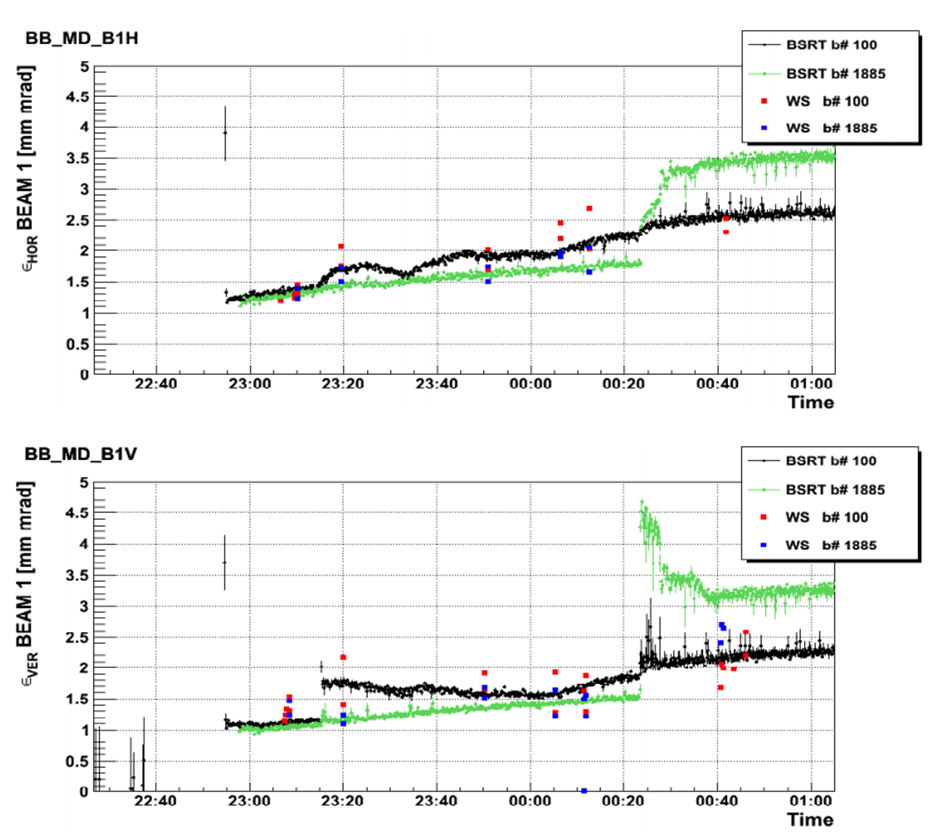}
\caption{Fill 1766: Beam 1 horizontal (upper plot) and vertical (lower plot) normalized emittance evolution for the bunches $B1-b100$ (black) and $B1-b991$ (green), as measured by the BSRT.}
\label{fig:BB_Wshop_1766_A_Emittance_B1}
\end{figure}

\begin{figure}[h!]
\centering
\includegraphics[width=.48\textwidth,angle=-00]{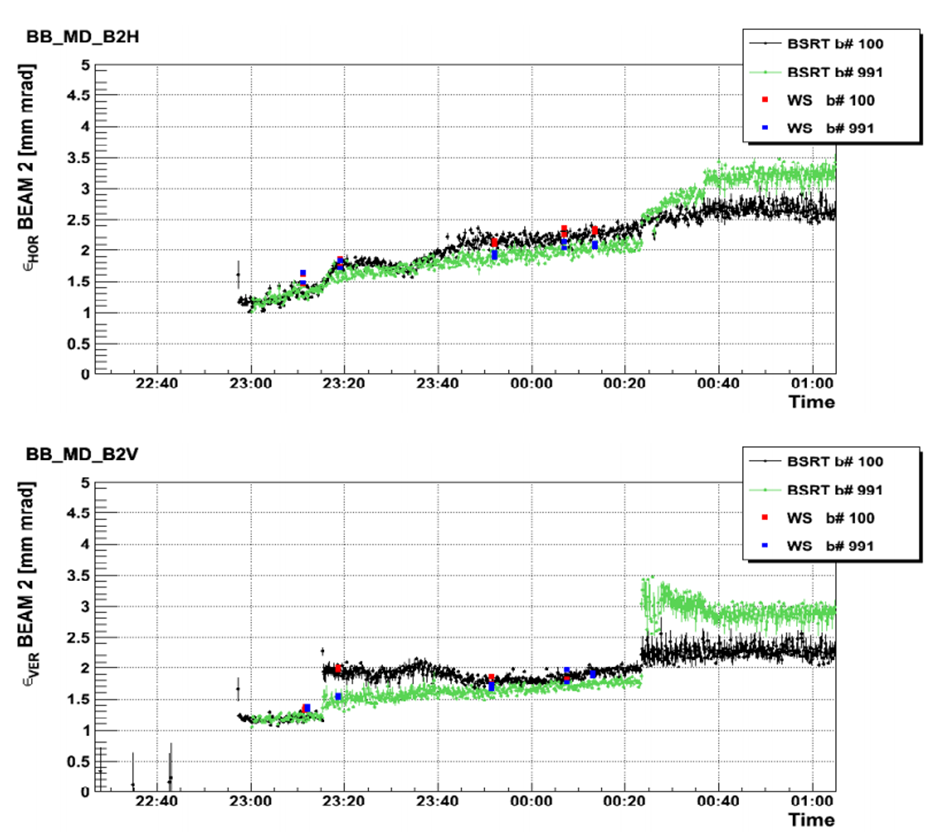}
\caption{Fill 1766: Beam 2 horizontal (upper plot) and vertical (lower plot) normalized emittance evolution for the bunches $B2-b100$ (black) and $B2-b991$ (green), as measured by the BSRT.}
\label{fig:BB_Wshop_1766_A_Emittance_B2}
\end{figure}

A signiﬁcant emittance increase was observed when beams started to collide in IP8. This emittance increase is observed on the bunches colliding only in IP8, i.e. with a single collision. The bunch with the largest number of collisions had the smallest emittance increase. The emittances as a function of time measured with the synchrotron light monitor (BSRT) are shown in Figs.~\ref{fig:BB_Wshop_1766_A_Emittance_B1} and~\ref{fig:BB_Wshop_1766_A_Emittance_B2}.
 
This emittance increase did not start when the bunches were put into collision, but during the optimization process, i.e. luminosity scan. It was done manually since the automatic procedure did not find an optimum.

\subsection{{Fill 1766: Part 2 (May 2011)}}
\begin{figure}[!h]
\centering
\includegraphics[width=.48\textwidth,angle=-00]{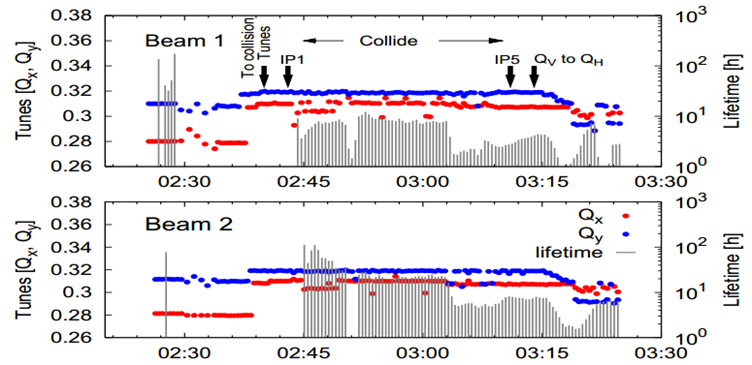}
\caption{Fill 1766: horizontal (blue) and vertical (red) tune and the lifetime (black) evolution  with the (0.31, 0.32) tune working point for Beam 1 (upper plot) and Beam 2 (lower plot).}
\label{fig:BB_Wshop_1766_B_OldTuneWP}
\end{figure}

\begin{figure}[]
\centering
\includegraphics[width=.48\textwidth,angle=-00]{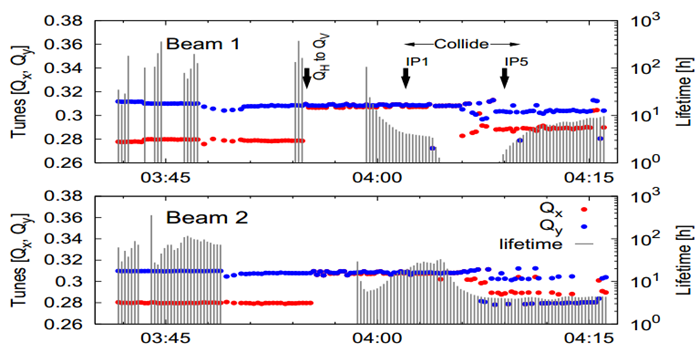}
\caption{Fill 1766: horizontal (blue) and vertical (red) tune and the lifetime (black) evolution  with the (0.31, 0.31) tune working point for Beam 1 (upper plot) and Beam 2 (lower plot).}
\label{fig:BB_Wshop_1766_B_NewTuneWP}
\end{figure}

With the same machine configuration as in Fill 1765, again two single bunches colliding only in IP1 and 5 were injected to study in more detail the parameter space, in particular the effect of the working point on the initial emittance growth observed once the beams are put into collision.
Two fillings were done using the same filling scheme with roughly the same injected intensities  ($\sim$1.8$\times$\,$10^{11}$\,ppb) and normalized emittance ($\sim$1.32\,$\mu$m), but changing the tunes starting working point.
As shown in Figs.~\ref{fig:BB_Wshop_1766_B_OldTuneWP} and~\ref{fig:BB_Wshop_1766_B_NewTuneWP}, it is seen that moving the tune point from (0.31, 0.32) to (0.31, 0.31) helped to increase the beams' lifetime.
The initial low lifetime observed was thought to originate from a lattice resonance where the core of the bunch was getting close to the 10th-order resonance; a small negative shift for the vertical tune ($\Delta Q_{v} = -0.01$) improved the losses and reduced the emittance blow-up since the large tune shift from the collisions will bring the core particles below the resonance.

Once this emittance blow-up was cured, a beam--beam linear parameter of $\xi=0.017$/IP was achieved resulting in a total value $\xi=0.034$.

\section{High pile-up tests: single bunches}
To provide a calibration test for the LHC experiments, tests were performed to generate a high pile-up in the detectors, 
i.e. a large number of events per crossing~\cite{MD_Note_2011_105, MD_Note_2012_010}.
For this purpose a large number of bunches is not needed and was done with single bunches per beam.
\subsection{Experimental Setup}
 For the LHC fill in study (Fill 2201), one high-intensity bunch ($\sim$2.4$\times$\,$10^{11}$\,ppb) was injected  in the bucket 1001 of both rings. The bunches were accelerated to the flat-top energy (3.5\,TeV) and put in collision in ATLAS and CMS according to the standard operational cycle, where the machine was in the standard configuration:\\
$\beta* = 1$\,m in IP1/5 and $\pm$120\,$\mu$rad crossing angle,\\
$\beta* = 10$\,m in IP2 and $\pm$80\,$\mu$rad crossing angle,\\
$\beta* = 3$\,m in IP8 and $\pm$250\,$\mu$rad crossing angle.\\

\subsection{Observations}
The resulting luminosity  was $ \sim 4.7 \times 10^{-30}$\,cm$^{-2}$\,s$^{-1} $ corresponding to a pile-up (number of inelastic interactions per crossing) of $ \mu\sim31 $ encounters per turn.
The nominal number is $ \mu\sim20 $~\cite{LHC_DESIGN_REPORT}.
Observations of the beam intensity and emittance  showed the following:

\subsubsection{Emittance}
The bunch transverse emittances were frequently measured with the WSs throughout the fill. At the injection energy an emittance growth was observed only in the horizontal plane (12\% for Beam 1 and 8\% for Beam 2). Through the ramp the beams' emittances in both planes steadily grew  reaching $\sim$3.1\,$\mu$m at the beginning of the  collisions (a growth of $ 44\%$ B1 and $ 27\% $ B2 in the horizontal plane and $ 30\%$ B1 and $ 7\%$ B2 in the vertical plane).
 In about 3\,h in collision the horizontal emittances grew by about 20\% for both beams while the vertical ones increased by $ \sim13\% $.\\

\subsubsection{Intensity}

\begin{figure}[]
\centering
\includegraphics[width=.475\textwidth,angle=-00]{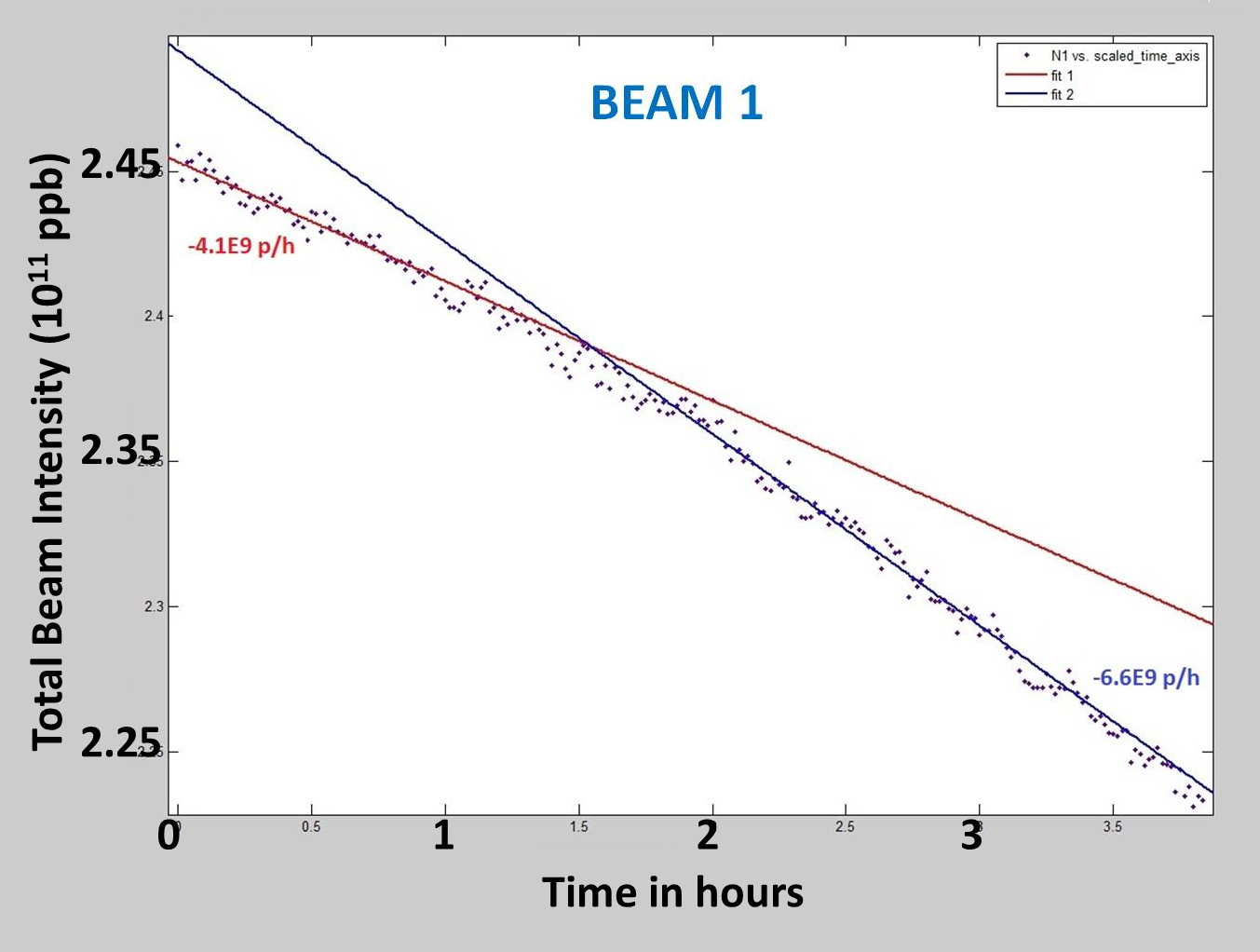}
\includegraphics[width=.48\textwidth,angle=-00]{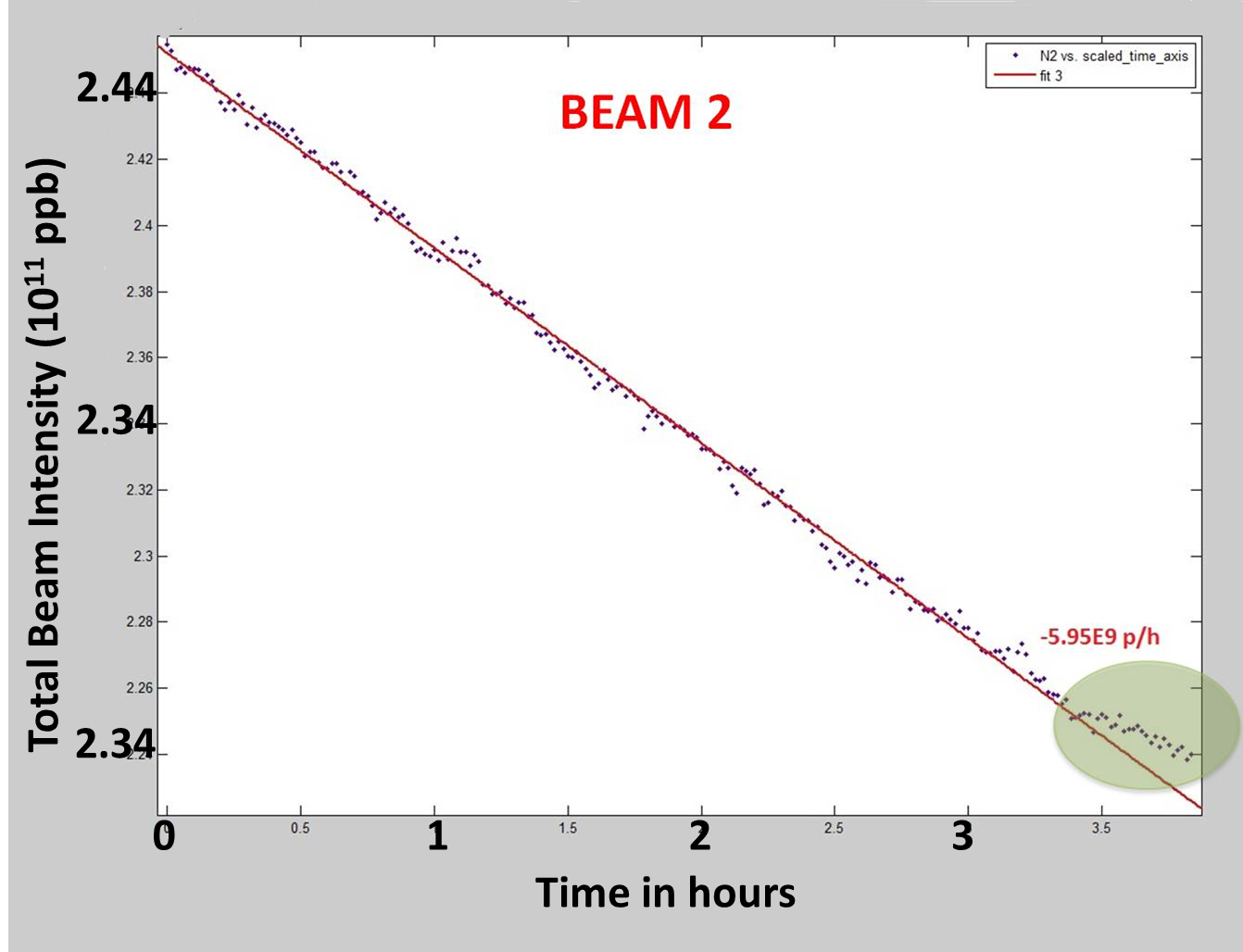}
\caption{Fill 2201: measured vs. fitted bunch intensity for Beam 1 (upper plot) and Beam 2 (lower plot). The points refer to the fBCT measurement and the lines are the linear fits representing the loss regimes observed. The slopes of the linear fits corresponding to the constant loss rate are also reported.}
\label{fig:BB_Wshop_2201_Intensity}
\end{figure}

The beam intensity decay was studied and two loss regimes were observed for Beam 1: the first regime ended after 1.5\,h in collision with a loss rate of $ \sim 4.1\times 10^{9}$\,p/h followed until the end of the fill with a constant loss regime of $ \sim6.6\times 10^{9}$\,p/h. For Beam 2, a constant loss rate was recorded throughout all the fill $ \sim 6\times 10^{9}$\,p/h. A correlation was observed between the transition between the two loss regimes in Beam 1 and its bunch length growth~\cite{LHC_BQM}, where, once the value of $ \sim 8.95$\,cm was reached, corresponding to the initial bunch length of Beam 2, the losses settled to their maximum value (Fig.~\ref{fig:BB_Wshop_2201_Intensity}).\\

\subsubsection{Luminosity}
 A validation of the measured beam parameters was made through a comparison between the published instantaneous luminosity from the experiments and the analytically calculated one. The observed discrepancy was compatible with the uncertainties in the values of the machine parameters ($ \beta^{*} $) and the monitors' accuracy.\\

In addition, the luminosity evolution model developed for the TEVATRON was applied using the LHC parameters, trying to predict the evolution of the beam parameters. An agreement was only found for the  period corresponding to the first loss regime for Beam 1, while for the second loss regime and for the Beam 2 intensity evolution, the longitudinal losses were underestimated almost by a factor of~3.

\section{2011 High-pile-up tests: multiple bunches}
After the single-bunch high-pile-up tests, the experimenters requested a new test to check whether multiple high-pile-up collisions can be effectively processed.

\subsection{Experimental Setup}
Fill 2252 was dedicated to providing ATLAS and CMS with high-pile-up collisions. A peak pile-up $\mu$ of almost 35 was achieved. Part of the test was also to bring the pile-up to values $< 0.5$ and this was achieved by separating the beams in steps.  The head-on collisions were restored after each IP separation to study its effects.
The collision scheme during this experiment is shown in Fig.~\ref{fig:BB_Wshop_2252_Collision_Scheme}.
The number of collisions is shown for every bunch.

The bunch parameters (intensity, emittance and length) of high-pile-up colliding bunches were observed while putting beams in collision with the same machine setup as in Fill 2201 but with the following filling scheme:
\begin{figure}[h!]
\includegraphics[width=.48\textwidth,angle=-00]{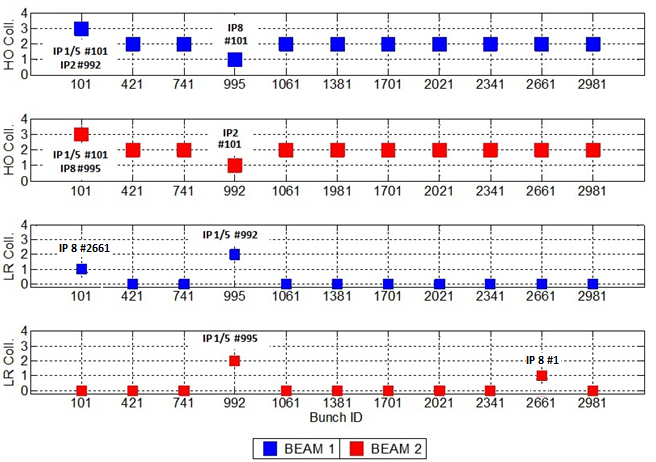}
\caption{Collision scheme of the injected bunches in Fill 2252. All bunches are colliding HO in IP1 and 5, except for $B1-b995$ and $B2-b992$ colliding respectively in IP8 with $B2-b101$  and in IP2 with $B1-b101$. The number of LR encounters is shown as well.}
\label{fig:BB_Wshop_2252_Collision_Scheme}
\end{figure}

\subsection{Observations}

After $\sim$\,100 min in single-bunch mode, the beams were re-separated into three different phases, as shown in Fig.~\ref{fig:BB_Wshop_2252_Luminosity_Separation_Steps}:

\begin{figure}[h!]
\includegraphics[width=.495\textwidth,angle=-00]{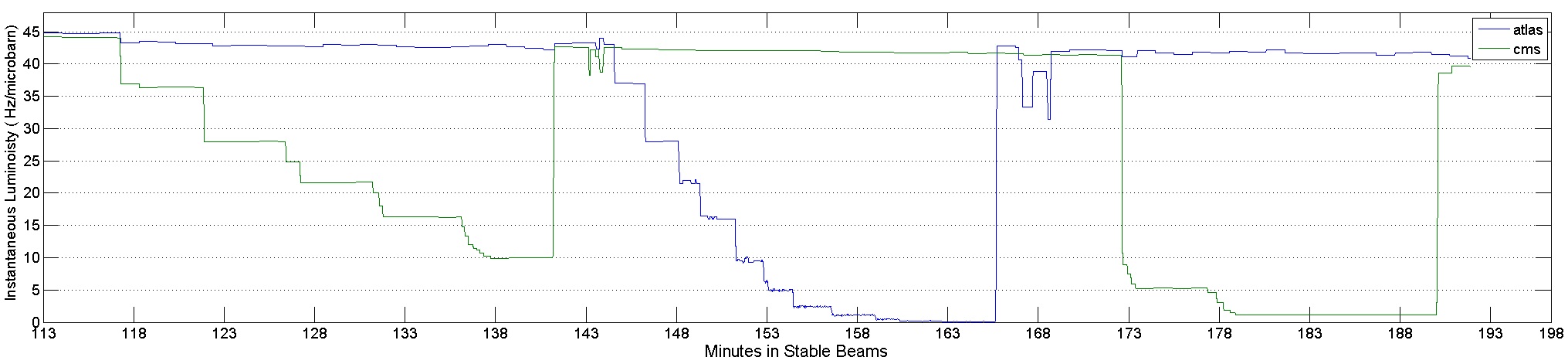}
\caption{ATLAS and CMS instantaneous luminosity during the separation phases shown respectively in blue and green.}
\label{fig:BB_Wshop_2252_Luminosity_Separation_Steps}
\end{figure}

\begin{itemize}
\item [--] Five separation steps in the horizontal plane of IP5 reducing the luminosity from $ \sim$\,44\,Hz/$\mu$b to $ \sim$\,10\,Hz/$\mu$b,
\item [--] Eleven separation steps in the vertical plane of IP1 reducing the luminosity from $ \sim$\,43\,Hz/$\mu$b to $ \sim$\,0.08\,Hz/$\mu$b,
\item [--] Two separation steps in the horizontal plane of IP5 reducing the luminosity from $ \sim$\,41\,Hz/$\mu$b to $ \sim$\,1.23\,Hz/$\mu$b.
\end{itemize}

After each separation, the instantaneous luminosity was re-optimized to its maximum value  in the considered IP.

\subsubsection{Emittance}
The bunches can be classified into three categories:

\begin{itemize}

\item [--] bunches presenting a higher growth rate in the vertical plane during the separation phases: $B2-b101$, $B2-b421$ and $B1-b1061$ (see Fig.~\ref{fig:BB_Wshop_2252_Emitt_Steady_Growth}),

\begin{figure}[h!]
\centering
\includegraphics[width=.48\textwidth,angle=-00]{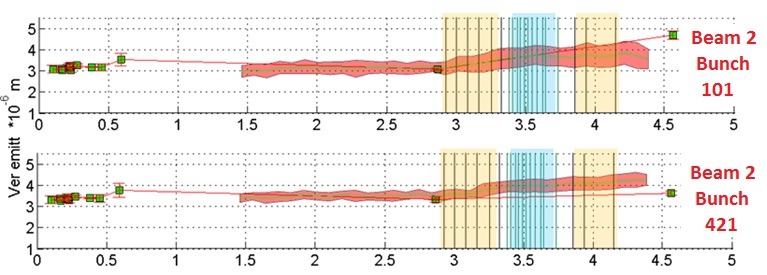}
\caption{Vertical emittances of bunches 101--421 (Beam 2) showing a continuous smooth increase in the emittance while separating the beams in  IP5. The vertical coloured bar delimits the separation steps of every separation phase in both IPs.}
\label{fig:BB_Wshop_2252_Emitt_Steady_Growth}
\end{figure}

\item [--] bunches presenting sudden blow-up throughout the separation steps: $B1-b101$, $B1-b421$, $B2-b992$ and $B2-b1061$ (see Fig.~\ref{fig:BB_Wshop_2252_Emitt_Sudden_Growth}).

\begin{figure}[h!]
\centering
\includegraphics[width=.48\textwidth,angle=-00]{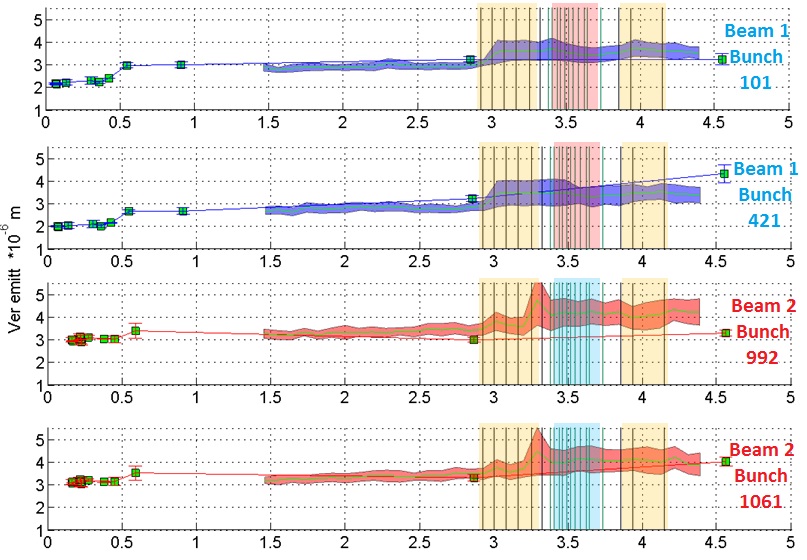}
\caption{Vertical blow-up of the vertical emittance of bunches 101--421 (Beam 1) and bunches 992--1061 (Beam 2) while separating the beams in IP5. The vertical coloured bar delimits the separation steps of every separation phase in both IPs.}
\label{fig:BB_Wshop_2252_Emitt_Sudden_Growth}
\end{figure}
\item [--] bunches not affected by the separation steps.

\end{itemize}

The sudden increases in emittance were observed mainly in the vertical plane during the steps of the first CMS separation (recall that in IP5 the vertical plane is the separation plane).

\subsubsection{Lifetime}
Normalizing the bunch intensity curves to the initial value at the start of collisions allows one to identify three groups of behaviours according to the number of HO collisions the bunches experience (three HO, two HO, one HO).

Analysing the intensity data for the first 1.8\,h, it is possible to disentangle the separation effects from the initial evolution of the bunch parameters.\\

\begin{figure}[h!]
\centering
\includegraphics[width=.48\textwidth,angle=-00]{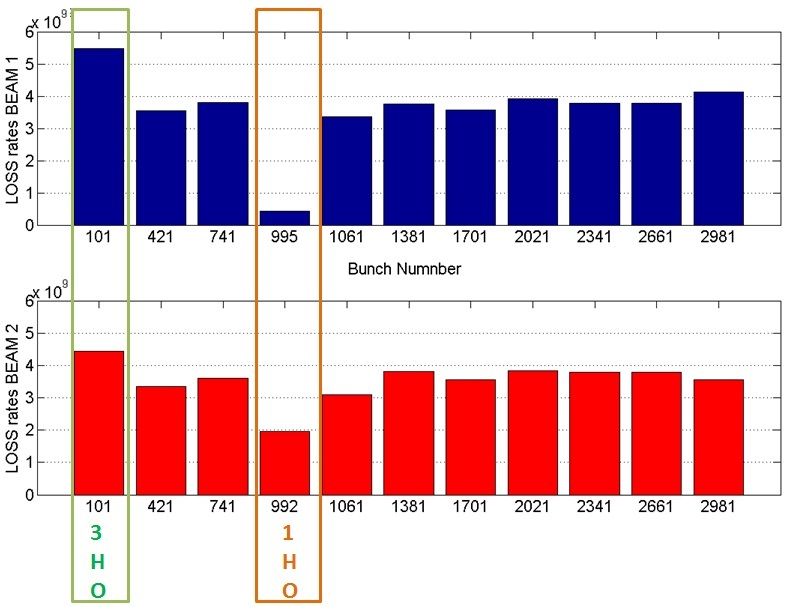}
\caption{Slopes from linear fits of the bunch intensity decays for the first 100\,min.}
\label{fig:BB_Wshop_2252_Intensity_Loss_Rates}
\end{figure}

Once in collision, for most bunches a linear decrease in intensity is observed during the considered time period; a linear fit is applied to the measured intensities and the slopes corresponding to the loss rates are shown in Fig.~\ref{fig:BB_Wshop_2252_Intensity_Loss_Rates}.

The highest losses are observed for bunches experiencing three HO collisions while the lowest losses are for bunches with one HO collision only.

We observe the losses for each of the three separation phases mentioned before; the loss trend can be organized into families (according to the collision scheme) for all the steps.

\begin{itemize}
\item [--] \emph{\textbf{Separation phase 1 (IP5)}}\\
The first separation step in IP5 had an effect on the luminosity in IP1, where a drop of 3.5\% was observed.
No significant change in the intensity evolution of all the bunches in both beams except for:
\begin{itemize}
\item [.] bunch 101 in Beam 1 (having three HO: IP1 and 5 with bunch 101 in Beam 2, IP2 with bunch 992 in Beam 2),
\item [.] bunch 992 in Beam 2 (having one HO: IP2 with bunch 101 in Beam 1).
\end{itemize}

An increase in the loss rate is observed for $B1-b101$ after the first separation step as shown in Fig.~\ref{fig:BB_Wshop_2252_Intensity_Loss_B2-b992}. For $B2-b992$, a sudden increase in losses is seen in the first and last steps of the separation. Note that at the end of the fifth step, once the beams are put back in HO collisions in IP5, the loss rate for $B2-b992$ returned to the value it had before the beams' manipulation as seen in Fig.~\ref{fig:BB_Wshop_2252_Intensity_Loss_B2-b992}.
In order to explain and understand the link between the behaviour of $B2-b992$, colliding only in IP2, and the separations in IP5 it is important to consider the LR collisions this bunch is experiencing in IP5 with $B1-b995B1 $.

\begin{figure}[h!]
\centering
\includegraphics[width=.48\textwidth,angle=-00]{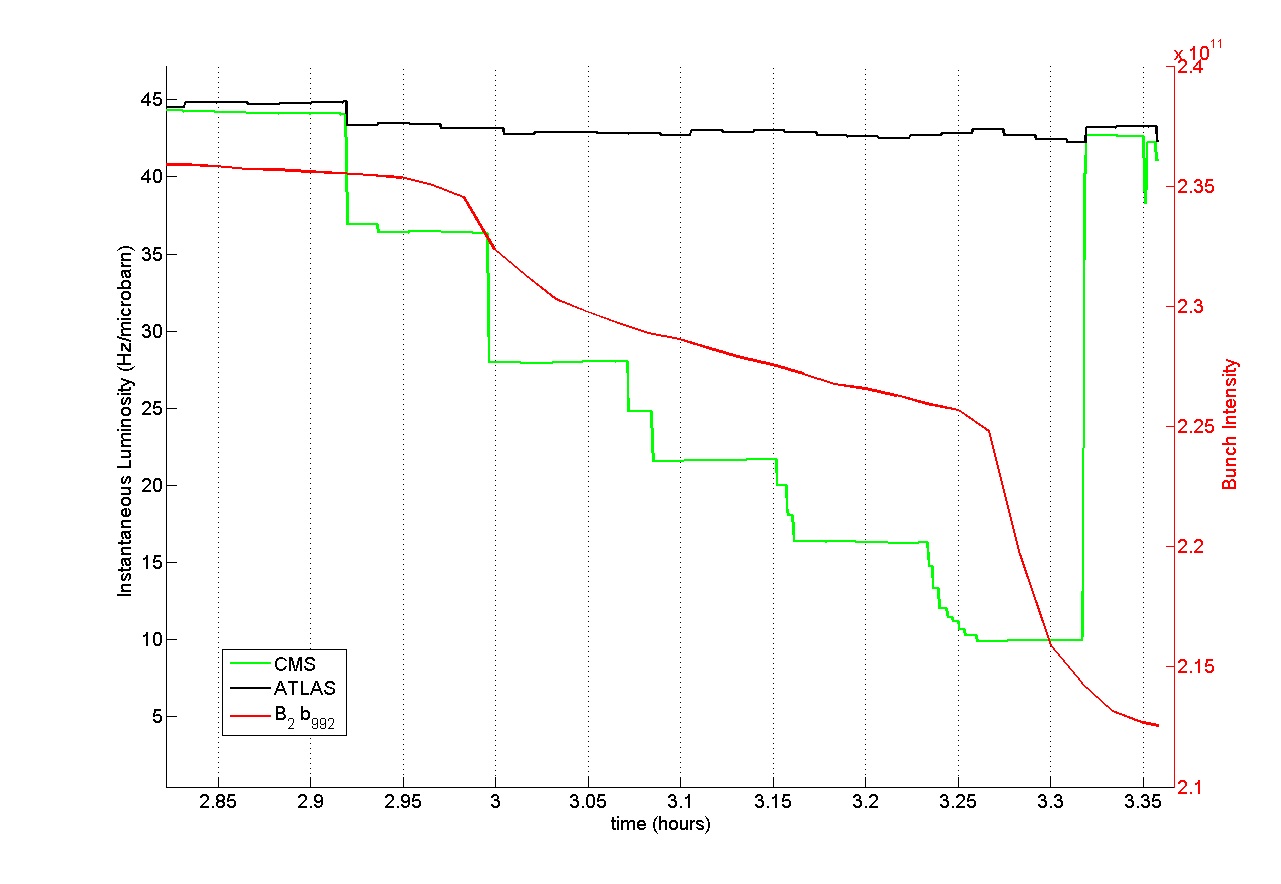}
\caption{The evolution of the normalized intensity of $B2-b992$ (red) during the first separation phase in IP5. The instantaneous luminosity in IP1 and IP5 is shown respectively in black and green.}
\label{fig:BB_Wshop_2252_Intensity_Loss_B2-b992}
\end{figure}

\item [--] \emph{\textbf{Separation phase 2 (IP1)}}\\
$B1-b995$ and $B2-b992$ having one HO collision in IP8 and 2, respectively, were not affected by the steps of this separation.
For the three HO collisions family, it can be observed that $B1-b101B1 $ experienced higher loss rate at half total separation and kept this loss regime until the beams were brought back into HO collision; bunch $ B2$-$b101 $ entered a higher loss regime after the first separation step and maintained this rate once back in the initial situation of fully HO collisions.
The slopes of the intensity evolution for all the other bunches (two HO families) depend on the value of the separation between the beams: lower losses were observed for higher separations. It is worth observing that the bunch losses for both families in Beam 1 and Beam 2 colliding in IP1 and 5 at zero separation were higher at the end of this phase (especially for Beam 1).

\item [--] \emph{\textbf{Separation phase 3 (IP5)}}\\
No clear variation in the intensity decay was observed for the bunches having one HO and three HO in both beams, while for all the other bunches, the losses were decreasing for an increasing beam separation. Again for some bunches in Beam 1 having two HO, the initial loss rate was not restored at the end when the beams were brought into HO collisions: higher losses were still observed at the end.
\end{itemize}

\subsubsection{Bunch length}
\begin{figure}[h!]
\centering
\includegraphics[width=.48\textwidth,angle=-00]{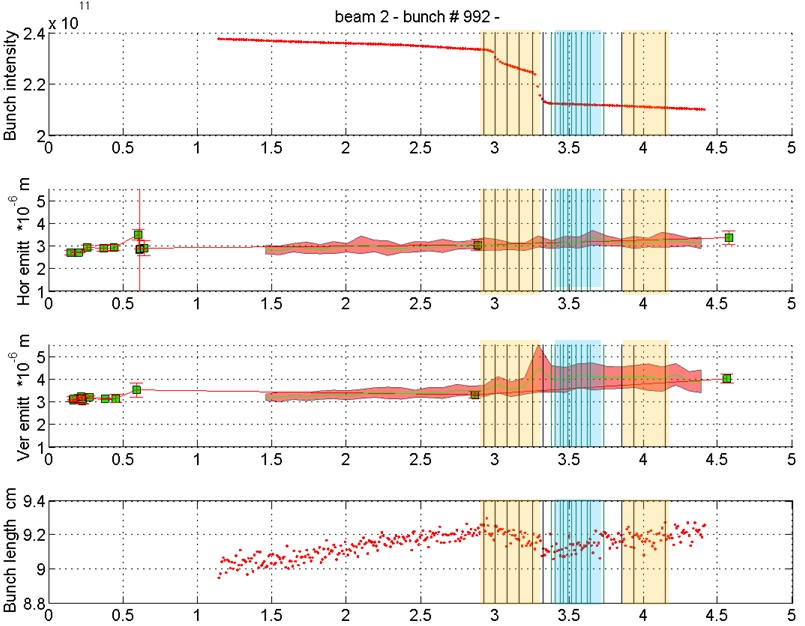}
\caption{The observed parameters for $B2-b992$ (intensity, transverse emittances and bunch length) from the start of the fill until the end of the collisions. The vertical coloured bar delimits the separation steps of every separation phase in both IPs. }
\label{fig:BB_Wshop_2252_AllParameters_B2-b992}
\end{figure}

We only highlight the bunch length evolution of Beam 2. It is worth pointing out the behaviour of $B2-b992$ (see Fig.~\ref{fig:BB_Wshop_2252_AllParameters_B2-b992}): a bunch shortening is observed simultaneously to the intensity loss, corresponding to the first separation step in  IP5 (first separation phase).
For all the other bunches, the bunch length did not bear signs of the losses and the emittance blow-up (horizontal and vertical).

\section{{2012 high-pile-up}}
\begin{figure}[h!]
\centering
\includegraphics[width=.48\textwidth,angle=-00]{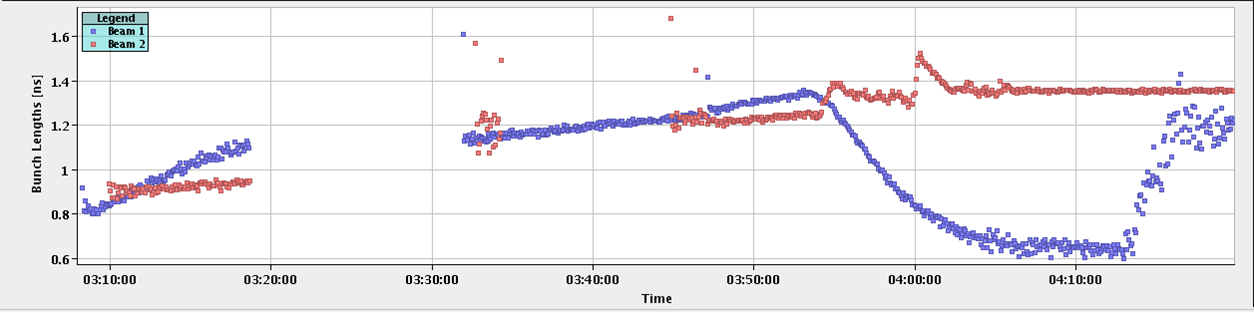}
\caption{Fill 2824: bunch length evolution for Beam 1 (blue) and Beam 2 (red) during the fill as measured by the beam-quality monitors in the LHC.}
\label{fig:BB_Wshop_2824_Long_BlowUP_Problem}
\end{figure}

\begin{figure}[h!]
\centering
\includegraphics[width=.5\textwidth,angle=-00]{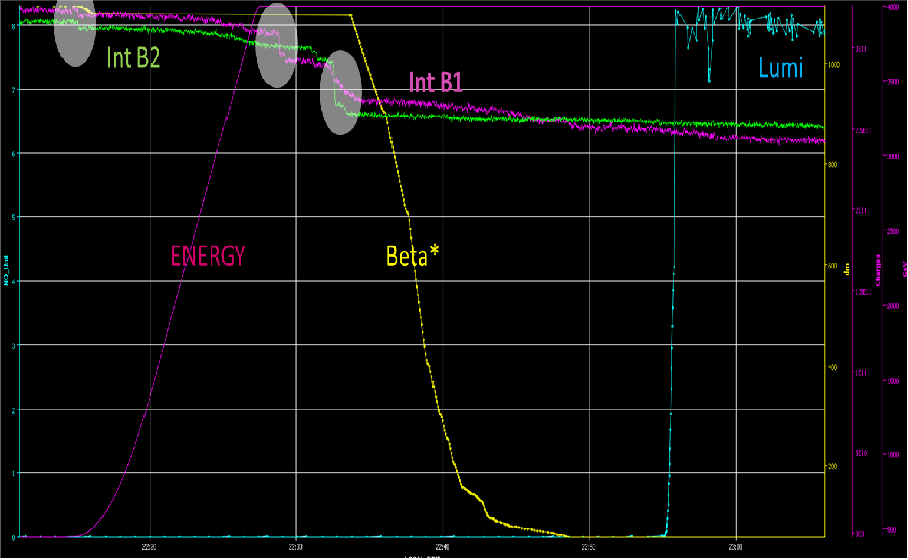}
\caption{Fill 2824: intensity evolution of Beam 1 (magenta) and Beam 2 (green) along with energy ramp (red). The value of $\beta^{*}$ in IP1 is shown as well to indicate the squeeze phase, and the instantaneous luminosity (blue) in IP1 is shown as well as an indication for the start of collisions.}
\label{fig:BB_Wshop_2824_Deteriorated_Beams}
\end{figure}

Profiting from the new optics in the SPS (Q20 optics), where the fractional part of the tune was moved to 0.2, very high brightness bunches were put into collision in the LHC aiming to establish a new record pile-up possibly up to 100.
High-brightness bunches with intensity of 3\,$\times$\,$10^{11}$\,ppb and normalized emittance of 2\,$\mu$m were used for this test.
The energy ramp was troublesome; the controlled longitudinal blow-up needed for the beam stability was faulty, as shown in Fig.~\ref{fig:BB_Wshop_2824_Long_BlowUP_Problem}. The bunch length of Beam 2 was brought to the target bunch length of 1.3\,ns while Beam 1 bunch length went down to $\sim 0.4$\,ns before reaching a value close to its target value of 1.2\,ns. This caused a deterioration in the beam quality throughout the ramp.
The bunches were put into collision with more than 10\% of losses (end of ramp and at flat-top energy) and a doubled emittance, as shown in Fig.~\ref{fig:BB_Wshop_2824_Deteriorated_Beams}.

The resulting pile-up of 58 was the best reached in two tries.
Once the issue with the longitudinal blow-up was solved, two bunches per beam colliding only in IP1 and IP5 were injected and with difficulty brought into collision at 4\,TeV with a $\beta^{*}$ of 60\,cm in both IPs. An instability arising during the squeeze was observed leading again to some losses along with an emittance increase. The maximum pile-up obtained was 70 in IP1 and 65 in IP5.

\section{Summary}
In this paper, we reported the main studies of the HO beam--beam effects in the LHC with nominal, high-intensity, high-brightness single and multiple bunches. It is worth noting that small contributions of the lattice non-linearities as well as good settings of the machine allowed one to quickly reach the nominal head-on beam--beam tune shift; it was also shown that the LHC allows very large head-on tune shifts above nominal.\\
It has to be seen whether this can be maintained in the presence of many LR interactions. Yet there is no reason to assume that an HO limit has been reached for the moment in the LHC.

\section{ACKNOWLEDGMENTS}
I would like to thank all the CERN beam--beam working group members for their contribution, help and advice, especially W.\,Herr and G.\,Papotti.

\end{document}